\documentclass[aps,prd,preprint,groupedaddress]{revtex4}
\usepackage{amssymb}
\usepackage{graphicx}
\usepackage{bm}
\usepackage{amsmath}

\begin{document}

%\preprint{}

%Title of paper
%%%%%%%%%%%%%%%%%%%%%%%%%%%%%%%%%%%%%%%%%%%%%%%
%%%%%%%%%%%%%%%%%%%%%%%%%%%%%%%%%%%%%%%%%%%%%%%
\title{Quantization of  massive scalar fields \\
over axis symmetric space-time backgrounds}
%%%%%%%%%%%%%%%%%%%%%%%%%%%%%%%%%%%%%%%%%%%%%%%
%%%%%%%%%%%%%%%%%%%%%%%%%%%%%%%%%%%%%%%%%%%%%%%
\author{Owen Pavel Fern\'{a}ndez Piedra}
%%%%%%%%%%%%%%%%%%%%%%%%%%%%%%%%%%%%%%%%%%%%%%%
\email{opavel@ucf.edu.cu }
%%%%%%%%%%%%%%%%%%%%%%%%%%%%%%%%%%%%%%%%%%%%%%%
\affiliation{Departamento de F\'{i}sica y Qu\'{i}mica, Universidad de Cienfuegos, Cuba}
%%%%%%%%%%%%%%%%%%%%%%%%%%%%%%%%%%%%%%%%%%%%%%%
%%%%%%%%%%%%%%%%%%%%%%%%%%%%%%%%%%%%%%%%%%%%%%%

\author{Alejandro Cabo Montes de Oca}
\email{cabo@icmf.inf.cu}
\affiliation{Grupo de F\'{i}sica Te\'{o}rica, ICIMAF, Cuba.}

\date{\today}

\begin{abstract}

 \noindent The renormalized mean value of the quantum  Lagrangian and
the Energy-Momentum tensor for  scalar fields coupled  to an
arbitrary gravitational field configuration are analytically
evaluated in the Schwinger-DeWitt approximation, up to second
order in the inverse mass value. The  cylindrical symmetry
situation  is considered. The results furnish the starting point
for investigating  iterative solutions of the back-reaction
problem related with the quantization of cylindrical scalar field
configurations. Due to the homogeneity of the  equations of motion
of the Klein-Gordon field, the general results are  also valid for
performing the quantization over either vanishing or non-vanishing
mean field configurations.  As an application, compact analytical
expressions are derived here for the quantum mean Lagrangian and
Energy-Momentum tensor in the particular background given by the
Black-String space-time.
\end{abstract}

% insert suggested PACS numbers in braces on next line
% insert suggested keywords - APS authors don't need to do this
%\keywords{}

\maketitle

%%%%%%%%%%%%%%%%%%%%%%%%%%%%%%%%%%%%%%%%%%%%%%
%\section{\label{sec:Intro} Introduction}
%%%%%%%%%%%%%%%%%%%%%%%%%%%%%%%%%%%%%%%%%%%%%%
\section{Introduction}

Semiclassical gravity considers the quantum dynamics of fields in a gravitational background, which  at this level of description is considered as a classical external field. That is, all fields are
considered as quantum ones, with the only exception of the external gravitational field, that remains satisfying the classical Einstein field equations, associated with sources given by the vacuum
expectation values of the stress energy tensor of the matter fields \cite{birrel,fulling}. In that situation, it is necessary to have adequate mathematical methods  to obtain explicit analytical
expressions for the renormalized stress tensor \(\langle T_{\mu}^{\nu}\rangle_{ren}\), the quantity that enters as a source in the semiclassical Einstein equations \cite{anderson,frolov, DeWitt,
avramidi,barvinsky,gilkey, matyjasek}. This stress tensor and the expectation value \(\langle\varphi^{2}\rangle_{ren}\) of a quantum field \(\varphi\) are the main objects to calculate from quantum
field theory in curved spacetime.

Having the components for \(\langle T_{\mu}^{\nu}\rangle_{ren}\), the backreaction of the quantized fields in the space-time geometry of black holes can in principle be determined, unless the
(unknown) effects of quantum gravity become important. For the calculation of \(\langle T_{\mu}^{\nu}\rangle_{ren}\) we need, in principle, the exact knowledge of the functional dependence of this
tensor over all the possible metrics. For this reason it is improbable to have an exact analytical formula for this object. Except for very special spacetimes , on which quantum matter fields
propagates, and for boundary conditions with a high degree of symmetry \cite{dowker,browncassidy,bunchdavies,allenfolacci,kirsten}, it is not possible to obtain exact expressions for this quantity.
That mathematical difficulty has led to the development of approximate methods to build the effective action, starting from which the energy momentum tensor can be calculated by functional
differentiation with respect to the metric. One of the developed techniques, the so called Schwinger-De Witt expansion, is based on a series development expansion of the effective action in inverse
powers of the field mass. It is well-known that this method can be used to investigate effects like the vacuum polarization of massive fields in curved backgrounds, whenever the Compton's wavelenght
of the field is less than the characteristic radius of curvature \cite{frolov, DeWitt, avramidi,barvinsky, matyjasek}. Also numerical computations of \(\langle T_{\mu}^{\nu}\rangle_{ren}\) and
\(\langle\varphi^{2}\rangle_{ren}\) have been performed by a number of authors \cite{howardcandelas,candelas,fawcet,jensen12,ADL,anderson,demello}.

In General Relativity there exists a four parameter family of black hole solutions called the generalized Kerr-Newmann family. The solutions belonging to this family are characterized by the four
parameters:  mass \(M\), angular momentum \(J\), charge \(Q\) and the Cosmological Constant \(\Lambda\) \cite{lemos}. These are axis-symmetric solutions that show different asymptotic behavior
depending on the sign of the cosmological constant. There are two important cases of axial symmetry. One is the spherical symmetry that have been studied in great detail since the birth of General
Relativity. The other one is the cylindrical symmetry. As it has been shown by Lemos in Ref. [\onlinecite{lemos}], in the case of negative cosmological constant, there exists a Black hole solution
showing cylindrical symmetry: the  so called Black String. Charged rotating black string solutions has many similarities with the Kerr-Newman black hole, apart from space-time being asymptotically
anti-de Sitter in the radial direction (and not asymptotically flat). The existence of black strings suggests that they could be the final state of  the collapse of matter having cylindrical
symmetry.

The problems of determining \(\langle\varphi^{2}\rangle_{ren}\) and investigate the renormalized stress tensor components for conformally coupled massless scalar fields in black String backgrounds
were studied by DeBenedictis in \cite{debenedictis1,debenedictis2}. Also the results obtained for \(\langle T_{\mu}^{\nu}\rangle_{ren}\) were used for the calculation of gravitational backreaction
of the quantum field. It was found that the perturbations initially strengthen the singularity, an effect similar to the case of spherical symmetry without a cosmological constant, indicating that
the behaviour of quantum effects may be universal and not depend on the geometry of the spacetime nor the presence of non-zero cosmological constant.

In this paper we address the problem of evaluating  the components of the renormalized vacuum expectation values of the Stress-Energy Tensor for a massive scalar field in a background space-time
having cylindrical symmetry.  The general results are applied to explicitly evaluate closed expression for those quantities in the special background formed by a  neutral and non-rotating
cylindrical Black String. In Section II  first we build the effective action and the Stress-Energy tensor taking into account terms up to the second order in the inverse mass  of the scalar field.
Section III is devoted to review the metric tensor  which solves the Einstein-Maxwell in the considered cylindric symmetry situation.  Finally, employing the  explicit form  of the Black-String
metric, close expressions for the renormalized components of the Energy-Momentum tensor are derived  in Section IV.   These  results can be used to study the vacuum polarization and the
back-reaction of the quantum scalar field in the gravitational background. Final comments and possible future
 extensions of the work are given in the section on Conclusions.

 In the following we use for the Riemann tensor, its contractions, and the covariant derivatives the sign conventions of Misner, Thorne and Wheeler \cite{misner}. Our units are such that
 \(\hbar=c=G=1\).

%%%%%%%%%%%%%%%%%%%%%%%%%%%%%%%%%%%%%%%%%%%%%
%\section{\label{sec:Intro} Introduction}
%%%%%%%%%%%%%%%%%%%%%%%%%%%%%%%%%%%%%%%%%%%%%%

%%%%%%%%%%%%%%%%%%%%%%

\section{Renormalized effective action}
This paper first consider the derivation of the Schwinger-De Witt approximation for the  renormalized Lagrangian and  Stress-Energy tensor of massive scalar field subject to an arbitrary background
space-time. Consider a single massive scalar field \(\phi(x)\) interacting with gravity with non minimal coupling constant \(\xi\) in four dimensions. The action for the system is:
\begin{equation}\label{}
    S=S_{gravity}+S_{matter}
\end{equation}
with \(S_{gravity}\) the Einstein-Hilbert action for the gravitational background field and \(S_{matter}\) that of the scalar field:
\begin{equation}\label{}
    S_{gravity}=\frac{1}{2}\int d^{4}x\sqrt{-g}\left(R-\Lambda\right)
\end{equation}
and:
\begin{equation}\label{}
    S_{matter}=-\frac{1}{2}\int d^{4}x\sqrt{-g}\left[g^{\mu\nu}\partial_{\mu}\phi\partial_{\nu}\phi+(m^{2}+\xi R)\phi^{2}\right]
\end{equation}
where $m$ is the mass of the field. The action (1) leads to the usual Einstein equations for the gravitational field and the Klein-Gordon one for the scalar field:
\begin{equation}\label{}
    R_{\mu\nu}-\frac{1}{2}g_{\mu\nu}R\,+\Lambda g_{\mu\nu}=8\pi\,T_{\mu\nu}
\end{equation}
\begin{equation}\label{}
    \left( \Box \,-\,\xi R\,-\,m^{2}\right )\phi\,=\,0,
\end{equation}
In the above equations \(\Box\,=\,g^{\mu\nu}\nabla_{\mu}\nabla_{\nu}\) is the covariant D'Alembertian and \(T_{\mu\nu}\) the stress tensor of the classical scalar field. Note that the values of the
scalar fields in the classical approximation can be non zero. However, in the particular case to be considered here in more detail, the Black String, the scalar field can be considered as vanishing.
The homogeneity of the scalar field equation, implies that the Black String solution  after fixing the classical scalar field  as vanishing, is also a solution of the classical Einstein equations
including scalar fields.

For a general background geometry, the most direct approach to evaluate the renormalized Lagrangian  is to use the first non-vanishing term of the renormalized effective action calculated using the
Schwinger -DeWitt approximation \cite{frolov}. The advantage of this approach lies in the purely geometric nature of the approximation that reflects its local nature. The effective action of the
quantized massive scalar field differs form the analogous actions constructed for fields of higher spins only by numerical coefficients, and one can generalize the presented results to fields of
other spins. However, it is important to stress that the method is restricted  to cases in which we avoid the presence of strong or rapidly varying gravitational fields. Moreover, nonphysical
divergences that appears in the massless limit obstacle its application in that case. The construction of the first-order renormalized stress-energy tensor should be carried out in an analytically
continued Euclidean space-time. The analytic continuation to the physical space is performed at the last stage of the calculations.

Using DeWitt's effective action approach and applying Schwinger's regularization prescription \cite{frolov, avramidi}  one gets the renormalized effective action for the quantized scalar field
satisfying equation (1) as
\begin{equation}\label{}
         W_{ren}\,=\,\int  d^{4}x \sqrt{-g}\,\mathfrak{L}_{ren}
\end{equation}
where the renormalized effective Lagrangian reads:
\begin{equation}
\mathfrak{L}_{ren}\,=\,{1\over 2(4\pi)^{2}\,} \sum_{k=3}^{\infty}{\,str \,a_{k}(x,x)\over k(k-1)(k-2)m^{2(n-2)}},
\end{equation}
and :
\begin{equation}\label{}
    str F=\left(-1^{i}\right)F^{i}_{i}=\int d^{4}x\left(-1^{A}\right)F^{A}_{A}(x)
\end{equation}
is the functional supertrace \cite{avramidi} . The coefficients \([a_{k}]= \,a_{k}(x,x')\), whose coincidence limit appears under the supertrace operation in (7) are the
Hadamard-Minakshisundaram-DeWitt-Seeley coefficients (HMDS), whose complexity rapidly increases with \(k\). As usual, the first three coefficients of the DeWitt-Schwinger expansion,
$a_{0},\,a_{1},\,{\rm and}\,a_{2}, $ contribute to the divergent part of the action and can be absorbed in the classical gravitational action by renormalization of the bare gravitational and
cosmological constants. Various authors have calculated some of the HDSM coefficients in exact form up to $n \geq 4$. DeWitt \cite{DeWitt} have calculated the coefficient $[a_{2}],$ which is
proportional to the trace anomaly of the renormalized Stress-Energy tensor of the quantized, massless, and conformally invariant fields. The coincidence limit of the coefficient $a_{3}$ has been
obtained by Gilkey \cite{gilkey}  whereas the coefficient $[a_{4}]$ has been calculated by Avramidi \cite{avramidi} .

Restricting ourselves here to the terms  proportional to $m^{-2},$ using integration by parts and the elementary properties of the Riemann tensor, we obtain for the renormalized effective lagrangian
,
\begin{equation}\label{}
    \mathfrak{L}_{ren}\,=\,\mathfrak{L}_{ren}^{conformal}+\widetilde{\mathfrak{L}_{ren}},
\end{equation}
where the conformal part of the effective lagrangian is given by
\begin{eqnarray}
 \nonumber
 \mathfrak{L}_{ren}^{conformal}\,&=&\,{1\over 192 \pi^{2} m^{2}} \left[ \Theta R
 \Box R\,+\,{1\over 140} R_{\mu \nu} \Box R^{\mu \nu}\right. \nonumber \\
 &-& {8\over 945} R^{\mu}_{\nu} R^{\nu}_{\gamma} R^{\gamma}_{\mu}
 \,+\,{2\over 315} R^{\mu \nu}
 R_{\gamma \varrho} R^{\gamma ~ \varrho}_{~ \mu ~ \nu}
\,+\, \,{1\over 1260} R_{\mu \nu} R^{\mu}_{~ \sigma \gamma \varrho} R^{\nu \sigma \gamma \varrho}
\nonumber \\
&+& \left.{17\over 7560} {R_{\gamma \varrho}}^{\mu \nu} {R_{\mu \nu}}^{\sigma \tau} {R_{ \sigma \tau}}^{\gamma \varrho} \,
 \,-\,{1\over 270}
R^{\gamma ~ \varrho}_{~ \mu ~ \nu}
 R^{\mu ~ \nu}_{~ \sigma ~ \tau} R^{\sigma ~ \tau}_{~ \gamma ~ \varrho}\right],
\end{eqnarray}
and the mass dependent contribution takes the form
\begin{eqnarray}
 \nonumber
 \widetilde{\mathfrak{L}_{ren}}\,&=&\,{1\over 192 \pi^{2} m^{2}} \left[ {1\over 30} \eta\left(R R_{\mu \nu } R^{\mu \nu}-R R_{\mu \nu \gamma \varrho} R^{\mu \nu \gamma \varrho}\right)\right. \nonumber \\
&+& \left. {1\over 2}\eta^{2} R \Box R
 \,-\,\eta^{3} R^{3}\right].
\end{eqnarray}
where we use \(\Theta={1\over 252} - {1\over 30} \xi\) and \(\eta=\xi-\frac{1}{6}\).

By standard functional differentiation of the effective action with respect to the metric,  the renormalized Stress-Energy tensor is obtained according to the known formula:
\begin{equation}\label{}
    \langle T_{\mu\nu}\rangle_{ren}=\frac{2}{\sqrt{\ -g}}\frac{\delta W_{ren}}{\delta g^{\mu\nu}}
\end{equation}
The result can be written in a general form as
\begin{equation}
   \langle T_{\mu}^{\ \ \nu}\rangle_{ren}=C_{\mu}^{\ \ \nu}+D_{\mu}^{\ \ \nu},
   \label{emTensor}
\end{equation}
where the $C_{\mu}^{\ \ \nu}$ and $D_{\mu}^{\ \ \nu}$ tensors take the somewhat cumbersome forms
\begin{eqnarray*}
 C_{\mu}^{\ \ \nu} &=&\frac{1}{96\pi^{2} m^{2}}\left[\Theta ( \nabla_{\mu}R\nabla^{\nu}R\,+\,\nabla^{\nu}\nabla_{\mu}(\Box
R)\,+\,\nabla_{\mu}\nabla^{\nu}(\Box R)\,-\,2 \Box^{2} R \delta_{\mu}^{\ \ \nu}\right.
\nonumber \\
&&\left. \,-\ {1\over 2} \delta_{\mu}^{\ \ \nu}\nabla_{\gamma}R\nabla^{\gamma}R \,-\,2 \Box R \nabla^{\nu}\nabla_{\mu}R )\,+\,{1\over 140}\left[\nabla_{\mu}R_{\gamma \lambda}\nabla^{\nu}R^{\gamma
\lambda}\,-\, \nabla^{\nu}R_{\gamma \lambda}\nabla^{\lambda}R^{\ \ \gamma}_{\mu}\right.\right.
\nonumber \\
&&\left.\left.\,-\, \nabla_{\mu}R_{\gamma \lambda}\nabla^{\lambda}R^{\gamma \nu}\,+ \nabla^{\gamma}R_{\gamma \lambda}\nabla^{\nu}R^{\ \ \lambda}_{\mu} \,+\,\nabla^{\gamma}R_{\gamma
\lambda}\nabla_{\mu}R^{\lambda \nu}\,+\, \nabla^{\gamma}\nabla^{\nu}(\Box R_{\gamma \mu})\,-\,\Box^{2} R_{\mu}^{\ \ \nu} \right.\right.
\nonumber \\
&&\left.\left.\,+\nabla^{\gamma}\nabla_{\mu}(\Box R_{\gamma}^{\ \nu}) \,-\, {1\over 2} \nabla_{\varrho}R_{\gamma \lambda}\nabla^{\varrho} R^{\gamma \lambda}\delta_{\mu}^{\ \nu}\,-
\nabla^{\gamma}\nabla^{\lambda}(\Box R_{\gamma \lambda})\delta_{\mu}^{\ \nu}\,+\,\nabla_{\lambda}\nabla^{\nu}R_{\gamma \mu}R^{\gamma \lambda} \right.\right.
\nonumber \\
&&\left.\left.\,+\, \nabla_{\lambda}\nabla_{\mu}R_{\gamma}^{\ \nu} R^{\gamma \lambda}\,-\ \nabla^{\gamma}\nabla^{\nu}R_{\gamma \lambda}R^{\ \ \gamma}_{\mu}\,-\,
 \left(\nabla_{\lambda} \nabla_{\sigma} R_{\ \gamma}^{\lambda \ \ \sigma \nu}\,+\,{1\over 2} \nabla^{\nu} \nabla_{\gamma}R \,-\, R^{\lambda \sigma}R_{\gamma \ \lambda \sigma}^{\ \nu}\right)R^{\ \ \gamma}_{\mu}
\right.\right.
\nonumber \\
&&\left.\left.\,+\, R_{\gamma\lambda}R^{\gamma\lambda}R^{\ \ \gamma}_{\mu} \,-\, \Box R_{\gamma \mu}R^{\gamma \nu} \,- \nabla^{\lambda}\nabla_{\mu}R_{\gamma \lambda}R^{\gamma \nu}\right]\,-\,{8\over
945}\left[{3\over 2} \nabla^{\nu}R_{\gamma \lambda}\nabla^{\lambda} R^{ \ \ \gamma}_{\mu}\right.\right.
\nonumber \\
&&\left.\left. \,+\, {3\over 2} \nabla^{\gamma} R_{\gamma \lambda}\nabla_{\mu}R^{\lambda \nu}\,-\, {3\over 2} \nabla^{\gamma} R_{\gamma \lambda} \nabla^{\varrho} R_{\varrho}^{ \ \lambda }
\delta_{\mu}^{\ \nu}\,-\, {3\over 2} \nabla_{\varrho}R_{\gamma \lambda} \nabla^{\lambda} R^{ \gamma \varrho} \delta_{\mu}^{\ \nu}\,+\,{3\over 2} \nabla_{\lambda} \nabla^{\nu}R_{\gamma \mu}R^{\gamma
\lambda}\right.\right.
\nonumber \\
&&\left.\left.\,+\, {3\over 2} \nabla_{\lambda} \nabla_{\mu}R_{\gamma}^{\ \nu} R^{\gamma \lambda}\,-\, {3\over 2} \nabla^{\lambda} \nabla_{\varrho}R_{\gamma \lambda}R^{\gamma \varrho} \delta_{\mu}^{
\ \nu}\,+\, {3\over 2} \nabla^{\lambda} \nabla^{\nu}R_{\gamma \lambda} R^{ \ \ \gamma}_{\mu} \,-\, {3\over 2} \left(\nabla_{\lambda} \nabla_{\sigma} R_{\ \gamma}^{\lambda \ \ \sigma
\nu}\right.\right.\right.
\nonumber \\
&&\left.\left.\left.\,+\,{1\over 2} \nabla^{\nu} \nabla_{\gamma}R \,-\, R^{\lambda \sigma}R_{\gamma \ \lambda \sigma}^{\ \nu}\,+\, R_{\gamma\lambda}R^{\gamma\lambda}\right) R^{\ \
\gamma}_{\mu}\right.\right.
\nonumber \\
&&\left.\left.\,+\,{3\over 2} \nabla^{\lambda} \nabla_{\mu}R_{\gamma \lambda} R^{\gamma \nu}\,-\, {3\over 2} \Box R_{\gamma \mu} R^{\gamma \nu} \,-\, {3\over 2} \nabla_{\varrho}
\nabla^{\gamma}R_{\gamma \lambda} R^{\lambda \varrho} \delta_{\mu}^{ \ \nu}\,+\,R_{\gamma \lambda} R_{\varrho}^{\ \gamma} R^{\lambda \varrho}\delta_{\mu}^{\ \nu}\right.\right.
\nonumber \\
\end{eqnarray*}
\begin{eqnarray*}
&&\left.\left. \,-\, 3 R_{\gamma \lambda} R^{\ \ \gamma}_{\mu} R^{\lambda \nu}\,+\, {3\over 2} \nabla_{\mu}R_{\gamma \lambda} \nabla^{\lambda} R^{\gamma \nu} \,-\, 3 \nabla_{\gamma} R_{\gamma \mu}
\nabla^{\lambda} R^{\gamma \nu}\,+\, {3 \over 2} \nabla^{\gamma}R_{\gamma \lambda} \nabla^{\nu} R^{\ \ \lambda}_{\mu}\right]\right.
\nonumber \\
&&\left.\,+\,{2\over 315}\left(\nabla^{\gamma}R_{\gamma \mu}\nabla_{\lambda} R_{\lambda}^{\ \nu}\,+\,\nabla_{\lambda}R_{\gamma}^{\ \nu}\nabla^{\gamma} R^{\ \ \lambda}_{\mu}\,-\, 2
\nabla^{\gamma}R_{\gamma \lambda} \nabla^{\lambda} R_{\mu}^{\ \ \nu}\,-\,\nabla^{\nu} R_{\gamma \lambda} \nabla^{\varrho} R_{\varrho \ \ \mu}^{\ \gamma \lambda} \right.\right.
\nonumber \\
&&\left.\left.\,+\,\nabla_{\varrho}R_{\gamma \lambda} \nabla_{\mu}R^{\gamma \varrho \lambda \nu} \,+\, 2 \nabla_{\varrho}R_{\gamma \lambda} \nabla^{\sigma}R_{\sigma}^{\ \gamma \lambda \varrho }
\delta_{\mu}^{\ \nu}\,-\,\nabla_{\lambda} \nabla_{\gamma} R_{\mu}^{\ \ \nu}R^{\gamma \lambda}\,+\, \nabla^{\varrho} \nabla^{\nu}R_{\gamma \lambda \varrho \mu}R^{\gamma \lambda}\right.\right.
\nonumber \\
&&\left.\left.\,-\,\Box R_{\gamma \mu \lambda}^{\ \ \ \ \nu} R^{\gamma \lambda}\,+\,\nabla^{\varrho} \nabla_{\mu}R_{\gamma \ \lambda \varrho} ^{\ \nu} R^{\gamma \lambda}\,-\,\nabla^{\lambda}
\nabla^{\sigma}R_{\gamma \lambda \varrho \sigma} R^{\gamma \varrho} \delta_{\mu}^{\ \nu}\,+\, {1\over 2} \nabla^{\gamma} \nabla_{\lambda}R_{\gamma}^{\ \ \nu} R^{\ \ \lambda}_{\mu}\right.\right.
\nonumber \\
&&\left.\left.\,+\,{1\over 2} \nabla_{\lambda} \nabla^{\gamma}R_{\gamma}^{\ \nu}R^{ \ \ \lambda}_{\mu} \,+\, {1\over 2} \nabla^{\gamma} \nabla_{\lambda}R_{\gamma \mu} R^{\lambda \nu}\,+\, {1\over 2}
\nabla_{\lambda} \nabla^{\gamma}R_{\gamma \mu} R^{\lambda \nu}\,+\, {1\over 2} R_{\gamma \lambda} R_{\varrho \sigma} R^{\gamma \varrho \lambda \sigma} \delta_{\mu}^{\ \nu} \right.\right.
\nonumber \\
&&\left.\left.\,-\,{3\over 2} R_{\gamma \lambda} R_{\varrho}^{\ \nu} R^{\gamma \varrho \lambda}_{\ \ \ \mu}\,-\, {3\over 2} R_{\gamma \lambda} R_{\varrho \mu} R^{\gamma \varrho \lambda
\nu}\,-\,\nabla^{\gamma} \nabla^{\lambda}R_{\gamma \lambda} R_{\mu}^{\ \ \nu} \,+\, \nabla_{\varrho} \nabla^{\nu}R_{\gamma \lambda} R^{\gamma \varrho \lambda}_{\ \ \ \mu}\right.\right.
\nonumber \\
&&\left.\left.\,+\, \nabla_{\varrho} \nabla_{\mu}R_{\gamma \lambda} R^{\gamma \varrho \lambda \nu}\,-\,\nabla_{\sigma} \nabla_{\varrho }R_{\gamma \lambda} R^{\gamma \sigma \lambda \varrho}
\delta_{\mu}^{\ \nu}\,-\, \Box R_{\gamma \lambda} R^{\gamma \ \ \lambda \nu}_{\ \mu}\,-\, \nabla_{\mu} R_{\gamma \lambda} \nabla^{\varrho} R_{\varrho}^{\ \gamma \lambda \nu}\right.\right.
\nonumber \\
&&\left.\left.\,+\,\nabla_{\varrho}R_{\gamma \lambda} \nabla^{\nu} R^{\ \ \gamma \varrho \lambda }_{\mu}\,-\,2 \nabla_{\varrho} R_{\gamma \lambda} \nabla^{\varrho}R^{\gamma \ \ \lambda \nu}_{\
\mu}\right)\,-\,{17\over 7560}\left(- 6 \nabla^{\varrho} R_{\gamma \lambda \varrho}^{\ \ \ \nu} \nabla^{\sigma} R_{\sigma \mu}^{\ \ \ \gamma \lambda }\right.\right.
\nonumber \\
&&\left.\left.\,+\,3 \nabla^{\varrho} \nabla_{\sigma} R_{\gamma \lambda \varrho}^{\ \ \ \nu} R^{\gamma \lambda \sigma}_{ \ \ \ \mu} \,-\, 3 \nabla^{\varrho} \nabla_{\sigma} R_{\gamma \lambda \varrho
\mu} R^{\gamma \lambda \sigma \nu}\,-\, 3 \nabla_{\sigma} \nabla^{\gamma} R_{\gamma \ \lambda \varrho}^{\ \nu} R^{\lambda \varrho \sigma}_{ \ \ \ \mu}\,-\, 3 R_{\gamma \lambda \varrho \mu} R_{\sigma
\tau}^{\ \ \varrho \nu} R^{\gamma \lambda \sigma \tau}\right.\right.
\nonumber \\
&&\left.\left.\,-\,3 \nabla_{\sigma} \nabla^{\gamma} R_{\gamma \mu \lambda \varrho }R^{\lambda \varrho \sigma \nu}\,+\, {1\over 2} R_{\gamma \lambda \varrho \sigma} R_{\tau \chi}^{\ \ \gamma
\lambda} R^{\varrho \sigma \tau \chi} \delta_{\mu}^{\ \nu}\,-\, 6 \nabla_{\sigma} R_{\gamma \lambda
\varrho }^{\ \ \ \nu} \nabla^{\varrho}R^{\gamma \lambda \sigma}_{\ \ \ \mu}\right)\right. \\
%\end{eqnarray*}
%\begin{eqnarray}
 &&\left.+ {1\over 1260}\left(\,-\,{1\over 2} \Box R_{\gamma \mu \lambda \varrho} R^{\gamma \nu \lambda \varrho}\,+\,2 \nabla_{\lambda}R_{\gamma \mu} \nabla^{\varrho}R_{\varrho}^{\ \nu \gamma \lambda} \,-\, 2 \nabla^{\gamma}R_{\gamma \lambda} \nabla^{\varrho} R_{\varrho
\ \ \mu}^{\ \nu \lambda}\right.\right.
\nonumber \\
&&\left.\left.\,+\,\nabla^{\nu}R_{\gamma \lambda \varrho \sigma} \nabla^{\sigma}R^{\gamma \lambda \varrho}_{\ \ \ \mu}\,-\,\nabla^{\gamma}R_{\gamma \lambda \varrho \sigma} \nabla^{\nu} R^{\lambda \
\ \varrho \sigma}_{\ \mu} \,-\, \nabla_{\sigma}R_{\gamma \lambda \varrho \mu} \nabla^{\sigma} R^{\gamma \lambda \varrho \nu}\,-\,2 \nabla_{\varrho}R_{\gamma \lambda} \nabla^{\lambda} R^{\gamma \ \
\varrho \nu}_{ \ \mu}\right.\right.
\nonumber \\
&&\left.\left.\,-\, {1\over 2} \nabla^{\gamma} R_{\gamma \lambda \varrho \sigma} \nabla^{\tau} R_{\tau}^{\ \lambda \varrho \sigma} \delta_{\mu}^{\ \nu} \,-\,2 \nabla_{\varrho} \nabla^{\gamma}
R_{\gamma \ \lambda \mu}^{\ \nu} R^{\lambda \varrho} \,-\, 2 \nabla^{\gamma} \nabla^{\varrho} R_{\gamma \lambda \varrho}^{\ \ \ \nu} R^{ \ \ \lambda}_{\mu}\,+\,2 \nabla_{\varrho} \nabla_{\lambda}
R_{\gamma \mu} R^{\gamma \varrho \lambda \nu}\right.\right.
\nonumber \\
%\end{eqnarray*}
%\begin{eqnarray}
&&\left.\left.\,+\,\nabla_{\sigma} \nabla^{\nu} R_{\gamma \mu \lambda \varrho} R^{\gamma \sigma \lambda \varrho}\,+\, R_{\gamma \mu } R_{\lambda \varrho \sigma}^{\ \ \ \nu} R^{\gamma \sigma \lambda
\varrho} \,-\,{1\over 2} \nabla^{\lambda} \nabla_{\tau}R_{\gamma \lambda \varrho \sigma} R^{\gamma \tau \varrho \sigma} \delta_{\mu}^{\ \nu}-2 \nabla_{\lambda}R_{\gamma \mu}
\nabla^{\varrho}R_{\varrho}^{\ \gamma \lambda \nu} \right.\right.
\nonumber \\
&&\left.\left.\,-\, {1\over 2}\nabla_{\tau} R_{\gamma \lambda \varrho \sigma} \nabla^{\sigma} R^{\gamma \lambda \varrho \tau} \delta_{\mu}^{\ \nu}\,-\,2 \nabla^{\gamma} \nabla_{\varrho} R_{\gamma
\lambda} R^{\lambda \ \ \varrho \nu}_{\ \mu} \,+\, {1\over 2} \nabla_{\tau} \nabla^{\gamma} R_{\gamma \lambda \varrho \sigma} R^{\lambda \tau \varrho \sigma} \delta_{\mu}^{\ \nu}\,+\,
\nabla_{\lambda} \nabla^{\nu} R_{\gamma \lambda \varrho \sigma} R^{\gamma \ \ \varrho \sigma}_{\ \mu}\right.\right.
\nonumber \\
&&\left.\left.\,-\,{1\over 2} \Box R_{\gamma \ \lambda \varrho }^{\ \nu} R^{\gamma \ \ \lambda \varrho}_{\ \mu}\right)\,-\,{1\over 270}\left(3 \nabla^{\gamma} R_{\gamma \lambda \varrho \mu}
\nabla^{\sigma} R_{\sigma}^{\ \varrho \lambda \nu}\,+\, {1\over 2} R_{\gamma \lambda \varrho \sigma} R_{\tau \ \chi}^{\ \gamma \ \varrho} R^{\lambda \tau \sigma \chi}\delta_{\mu}^{\
\nu}\right.\right.
\nonumber \\
&&\left.\left.\,+\,3 \nabla^{\gamma} R_{\gamma \lambda \varrho \sigma} \nabla^{\sigma} R^{\gamma \nu \varrho}_{ \ \ \ \mu} \,-\,{3\over 2} \nabla_{\sigma} \nabla_{\varrho} R_{\gamma \mu \lambda}^{\
\ \ \ \nu} R^{\gamma \sigma \lambda \varrho}\,-\, {3\over 2} \nabla_{\sigma} \nabla_{\varrho} R_{\gamma \ \lambda \ \mu}^{\ \nu } R^{\gamma \sigma \lambda \varrho}\right.\right.
\nonumber \\
&&\left.\left.\,+\,{3\over 2} \nabla_{\sigma} \nabla^{\varrho} R_{\gamma \ \lambda \varrho }^{\ \nu} R^{\gamma \sigma \lambda }_{\ \ \ \mu}\,+\, {3\over 2} \nabla_{\sigma} \nabla^{\varrho} R_{\gamma
\mu \lambda \varrho} R^{\gamma \sigma \lambda \nu} \,-\, 3 R_{\gamma \lambda \varrho \mu} R_{\sigma \ \tau}^{\ \lambda \ \nu} R^{\gamma \sigma \varrho \tau}\right.\right.
\nonumber \\
&&\left.\left.\,+\,{3\over 2} \nabla_{\sigma} \nabla^{\lambda} R_{\gamma \lambda \varrho }^{\ \ \ \nu} R^{\gamma \ \  \varrho \sigma}_{ \ \mu}\,-\, {3\over 2} \nabla^{\lambda} \nabla^{\sigma}
R_{\gamma \lambda \varrho \sigma} R^{\gamma \ \ \varrho \nu}_{\ \mu}\,+\, {3\over 2} \nabla^{\lambda} \nabla_{\sigma} R_{\gamma \lambda \varrho \mu} R^{\gamma \nu \varrho \sigma}\right.\right.
\nonumber \\
&&\left.\left.\,+\, 3 \nabla^{\gamma} R_{\gamma \lambda \varrho \sigma} \nabla^{\sigma} R^{\lambda \ \ \varrho \nu}_{\ \mu}\,-\, {3 \over 2} \nabla^{\lambda} \nabla^{\sigma} R_{\gamma \lambda
\varrho \sigma} R^{\gamma \nu \varrho}_{\ \ \ \mu} \,+\, 3 \nabla_{\sigma} R_{\gamma \lambda \varrho }^{\ \ \ \nu} \nabla^{\lambda}R^{\gamma \ \ \varrho \sigma}_{\ \mu}\right)\right].
\end{eqnarray*}
and:
\begin{eqnarray*}
D_{\mu}^{\ \ \nu} &=&\frac{1}{96\pi^{2} m^{2}}\left[{1\over 30}\eta\left( \nabla^{\nu}R \nabla^{\gamma} R_{\gamma \mu}\,+\,\nabla_{\mu}R \nabla^{\gamma}R_{\gamma}^{\ \nu} \,+\,2
\nabla^{\nu}R_{\gamma \lambda}\nabla_{\mu} R^{\gamma \lambda}\,-\, \Box R R_{\mu}^{\ \ \nu}\right.\right.
\nonumber \\
&&\left.\left. \,+\, \nabla_{\gamma}R \nabla^{\nu}R^{\ \ \gamma}_{\mu} \,+\, \nabla_{\gamma}R \nabla_{\mu} R^{\gamma \nu} \,-\, 2 \nabla_{\gamma}R \nabla^{\gamma} R_{\mu}^{\ \ \nu}\,+\ R
\nabla^{\gamma}\nabla^{\nu}R_{\gamma \mu}\,+\,R \nabla^{\gamma}\nabla_{\mu}R_{\gamma}^{\ \nu} \right.\right.
\nonumber \\
&&\left.\left. -R \nabla_{\lambda} \nabla_{\gamma} R_{\ \mu}^{\lambda \ \ \gamma \nu}\,-\,{1\over 2}R \nabla^{\nu} \nabla_{\mu}R \,+\,R R^{\lambda \gamma}R_{\mu \lambda \ \gamma}^{\ \ \ \nu}\,-\,R
R_{\mu\lambda}R^{\mu\lambda}\,-\ 2 \nabla_{\gamma}R \nabla^{\lambda}R_{\lambda}^{\ \gamma} \delta_{\mu}^{\ \nu}\right.\right.
\nonumber \\
&&\left.\left. \,-\nabla^{\nu}\nabla_{\mu}R_{\gamma \lambda} R^{\gamma \lambda}\,+\,\nabla_{\mu}\nabla^{\nu}R_{\gamma \lambda} R^{\gamma \lambda} \,+\, \nabla_{\lambda}\nabla_{\gamma}R R^{\gamma
\lambda} \delta_{\mu}^{\ \nu}\,-\ 2 \Box R_{\gamma \lambda} R^{\gamma \lambda} \delta_{\mu}^{\ \nu} \right.\right.
\nonumber \\
&&\left.\left. \,+\, {1\over 2} R R_{\gamma \lambda} R^{\gamma \lambda} \delta_{\mu}^{\ \nu}\,+\,\nabla^{\nu}\nabla_{\gamma}R R^{\ \ \gamma}_{\mu}\,-\ 2 R R_{\gamma}^{\ \nu} R^{\ \ \gamma}_{\mu}
\,+\,\nabla_{\mu}\nabla_{\gamma}R R^{\gamma \nu}\,-\ R_{\gamma \lambda} R^{\gamma \lambda} R_{\mu}^{\ \ \nu}\right.\right.
\nonumber \\
&&\left.\left. \,+\,4 \nabla_{\gamma} R \nabla^{\lambda} R_{\lambda \mu}^{\ \ \ \gamma \nu}\,+\,2 \nabla_{\tau}R_{\gamma \lambda \varrho \sigma}\nabla^{\tau} R^{\gamma \lambda \varrho \sigma }
\delta_{\mu}^{\ \nu}\,+\,4 \nabla_{\gamma}R \nabla^{\lambda} R_{\lambda \ \ \mu}^{\ \nu \gamma}\,-\, 2 \nabla^{\nu}R_{\gamma \lambda \varrho \sigma}\nabla_{\mu}R^{\gamma \lambda \varrho \sigma
}\right.\right.
\nonumber \\
&&\left.\left.\,+\, 2 R \nabla^{\gamma}\nabla^{\lambda}R_{\gamma \mu \lambda}^{ \ \ \ \ \nu } \,- \nabla^{\nu} \nabla_{\mu}R_{\gamma \lambda \varrho \sigma}R^{ \gamma \lambda \varrho \sigma}\,-\,
\nabla_{\mu} \nabla^{\nu}R_{\gamma \lambda \varrho \sigma} R^{ \gamma \lambda \varrho \sigma} \,+\, 2 R \nabla^{\gamma} \nabla^{\lambda}R_{\gamma \ \lambda \mu}^{\ \nu} \right.\right.
\nonumber \\
&&\left.\left.\,+\, 2 \Box R_{\gamma \lambda \varrho \sigma} R^{\gamma \lambda \varrho \sigma}\delta_{\mu}^{\ \nu}\,- \ {1\over 2} R R_{\gamma \lambda \varrho \sigma} R^{\gamma \lambda \varrho
\sigma} \delta_{\mu}^{\ \nu}\,+\, R_{\mu}^{\ \ \nu} R_{\gamma \lambda \varrho \sigma} R^{\gamma \lambda \varrho \sigma}\,+\, 2 R R_{\gamma \lambda \varrho}^{\ \ \ \nu} R^{\gamma \lambda \varrho}_{\
\ \ \mu}\right.\right.
\nonumber \\
&&\left.\left.\,+\, 2 \nabla_{\lambda} \nabla_{\gamma}R R_{\ \mu}^{\gamma \ \ \lambda \nu}\,+\,2 \nabla_{\lambda} \nabla_{\gamma}R R^{\gamma \nu \lambda}_{\ \ \ \ \mu}\,-\, 2
\nabla_{\varrho}R_{\gamma \lambda}\nabla^{\varrho} R^{\gamma \lambda} \delta_{\mu}^{\ \nu}\,-\,R \nabla^{\gamma}\nabla^{\lambda}R_{\gamma \lambda}\delta_{\mu}^{\ \nu}\right)\right.
\nonumber \\
&&\left.\,+\,{1\over 2}\eta^{2}(\nabla_{\mu}R\nabla^{\nu}R\,+\,\nabla^{\nu}\nabla_{\mu}(\Box R)\,+\,\nabla_{\mu}\nabla^{\nu}(\Box R)\,-\ {1\over 2} \delta_{\mu}^{\ \
\nu}\nabla_{\gamma}R\nabla^{\gamma}R \right.
\nonumber \\
&&\left.\,-\,2 \Box^{2} R \delta_{\mu}^{\ \ \nu}\,-\,2 \Box R \nabla^{\nu}\nabla_{\mu}R)\,-\,\eta^{3}(6\nabla_{\mu}R\nabla^{\nu}R \,+\, 6 R \nabla^{\nu}\nabla_{\mu}R\,+\, {1\over 2} R^{3}
\delta_{\mu}^{\ \nu}\right.
\nonumber \\
&&\left.
 \,-\, 6 R \Box R \delta_{\mu}^{\ \nu}\,-\, 3 R^{2} R_{\mu}^{\ \
 \nu}\,-\,6 \nabla_{\gamma}R\nabla^{\gamma}R \delta_{\mu}^{\ \nu})\right], \\
\end{eqnarray*}

\section{The Black String metric}

 Let us review in this section the particular metric associated to the Black
String solutions. Consider  the  Einstein-Hilbert action in four dimensions with a cosmological term in the presence of an electromagnetic field. The total  action will be
\begin{equation}\label{}
    S=S_{gravity}+S_{em},
\end{equation}
where \(S_{gravity}\) is given by (2) and:
\begin{equation}\label{}
    S_{em}=-\frac{1}{2}\int
    d^{4}x\sqrt{-g}F^{\mu\nu}F_{\mu\nu}.
\end{equation}
The Maxwell tensor is
\begin{equation}\label{}
    F_{\mu\nu}=\partial_{\mu}A_{\nu}-\partial_{\nu}A_{\mu},
\end{equation}
\(A_{\mu}\) being the vector potential.  We will consider in this work solution of the Einstein-Maxwell system admitting a commutative two dimensional Lie group \(G_{2}\) of isometries. These are
space-times with cylindrical symmetry. The group \(G_{2}\) generates two dimensional spaces with three possible topologies. The first is \(R\times S^{1}\), the standard cylindrically symmetric
model, with orbits diffeomorphic either to cylinders or to R (i.e, \(G_{2}=R\times U(1)\)), the second one is \(S_{1}\times S_{1}\) the flat torus \(T_{2}\) model (\(G_{2}=U(1)\times U(1)\)), and
finally the third possible topology is \(R_{2}\). However, we will focus only on the first case. Then  a cylindrical coordinate system \((x^{0},x^{1},x^{2},x^{3})=(t,\rho,\varphi,z)\) with
\(-\infty<t<\infty\), \(0\leq \rho<\infty\), \(-\infty<z<\infty\), \(0\leq \varphi<\infty\) will be chosen. An stationary cylindrically symmetric spacetime that solves the Einstein-Maxwell equations
from (1) is  (see Ref. \cite{lemos}):
\begin{multline}\label{}
    \ \ \ \ \ \ \ \ \ \ \ \ \ ds^{2}=-(\alpha^{2}\rho^{2}-\frac{2(M+\Omega)}{\alpha\rho}+\frac{4Q^{2}}{\alpha^{2}\rho^{2}})dt^{2}-\frac{16J}{3\alpha\rho}(1-\frac{2Q^{2}}{(M+\Omega)\alpha\rho})dtd\varphi\\
    +[\rho^{2}+\frac{4(M-\Omega)}{\alpha^{3}\rho}(1-\frac{2Q^{2}}{(M+\Omega)\alpha\rho})]d\varphi^{2} \ \ \ \ \ \ \ \ \ \ \ \ \ \ \ \ \ \ \ \ \ \
    \\+\frac{1}{\alpha^{2}\rho^{2}-\frac{2(3\Omega-M)}{\alpha\rho}+\frac{4Q^{2}(3\Omega-M)}{\alpha^{2}\rho^{2}(\Omega+M)}}d\rho^{2}+\alpha^{2}\rho^{2}dz^{2} \  \label{ds2}\
\end{multline}
where \(M\), \(Q\), and \(J\) are the mass, charge, and angular
momentum per unit length of the string respectively. \(\Omega\) is
given by
\begin{equation}\label{}
    \Omega=\sqrt{M^{2}-\frac{8J^{2}\alpha^{2}}{9}}.
\end{equation}
The constant \(\alpha\) is defined as follows:
\begin{equation}\label{}
    \alpha^{2}=-\frac{1}{3}\Lambda,
\end{equation}
where \(\Lambda\) is a negative  Cosmological Constant, giving to
the spacetime its asymptotically anti-De Sitter behavior. In this
paper we are only concerned with space-times showing both charge
and angular momentum equal to zero, thus yielding the following
form to relation (\ref{ds2}):
\begin{equation}\label{}
    ds^{2}=-(\alpha^{2}\rho^{2}-\frac{4M}{\alpha\rho})dt^{2}+
    \frac{1}{(\alpha^{2}\rho^{2}-\frac{4M}{\alpha\rho})}d\rho^{2}+
    \rho^{2}d\varphi^{2}+\alpha^{2}\rho^{2}dz^{2}.  \label{ds2simp}
\end{equation}
As immediately can  be seen from (\ref{ds2simp}),  the considered
metric  behaves as the one corresponding to the  anti-De Sitter
space-time in the limit \(\rho\rightarrow\infty\),  and therefore
is not globally hyperbolic. This solution has an event horizon
located at \(\rho_{H}=\frac{\sqrt[3]{4M}}{\alpha}\) and the
apparent singular behavior at this horizon is a coordinate effect
and not a true one. The only true singularity is a polynomial one
at the origin, as can it be seen after  calculating the
Kretschmann scalar. It  results in
\begin{equation}\label{}
    K=R_{\alpha\beta\xi\gamma}R^{\alpha\beta\xi\gamma}=
    24\alpha^{4}\left(1+\frac{M^{2}}{\alpha^{6}\rho^{6}}\right).
\end{equation}

\section{Renormalized Stress-Energy tensor for scalar fields in a
Black String background}

In the space-time of a static Black String metric given by (\ref{ds2simp}) simple results were obtained for the renormalized Stress Tensor of massive scalar field showing an arbitrary coupling to
the background gravitational field. After a direct calculation, for the conformal part of the stress tensor we evaluated  in this work the result:
\begin{equation}\label{}
    C_{t}^{\ t}=\frac{1}{2520m^{2}\pi^{2}\alpha^{3}\rho^{9}}\left(11\alpha^{9}\rho^{9}
    -201\alpha ^{3}M^{2}\rho^{3}+1252M^{3}\right),
\end{equation}
\begin{equation}\label{}
    C_{z}^{\ z}=C_{\varphi}^{\ \varphi}=\frac{1}{2520m^{2}\pi^{2}\alpha^{3}\rho^{9}}
    \left(11\alpha^{9}\rho^{9}-183\alpha
    ^{3}M^{2}\rho^{3}+1468M^{3}\right),
\end{equation}
\begin{equation}\label{}
    C_{\rho}^{\ \rho}=\frac{1}{2520m^{2}\pi^{2}\alpha^{3}\rho^{9}}
    \left(11\alpha^{9}\rho^{9}+189\alpha
    ^{3}M^{2}\rho^{3}-308M^{3}\right).
\end{equation}
The above components  of the Stress Tensor do not depend in any way  of the coupling constant \(\xi\) because of the constant value of the Ricci scalar in this space-time:
\(R=-12\alpha^{2}=4\Lambda\). The coupling parameter arises only in the term proportional to the D\'Alembertian of the Ricci scalar that in our case is identically zero. The results for the
components of the $D_{\mu}^{\ \ \nu}$ tensor are
\begin{equation}\label{}
    D_{t}^{\ t}=\eta\left[\frac{1}{80\alpha^{3}\rho^{9}\pi^{2}m^{2}}\left(112\alpha^{3}\rho
    ^{3}M^{2}-704M^{3}+\alpha^{9}\rho^{9}\right)\right]-\frac{9\alpha^{6}}{2\pi^{2}m^{2}}\eta^{3},
\end{equation}
\begin{equation}\label{}
    D_{z}^{\ z}=D_{\varphi}^{\ \varphi}=\eta\left[\frac{1}{80\alpha^{3}\rho^{9}\pi^{2}m^{2}}\left(112\alpha^{3}\rho
    ^{3}M^{2}-896M^{3}+\alpha^{9}\rho^{9}\right)\right]-\frac{9\alpha^{6}}{2\pi^{2}m^{2}}\eta^{3},
\end{equation}
\begin{equation}\label{}
    D_{\rho}^{\ \rho}=\eta\left[\frac{1}{80\alpha^{3}\rho^{9}\pi^{2}m^{2}}\left(-112\alpha^{3}\rho
    ^{3}M^{2}+192M^{3}+\alpha^{9}\rho^{9}\right)\right]-\frac{9\alpha^{6}}{2\pi^{2}m^{2}}\eta^{3}.
\end{equation}
It is interesting to evaluate the above components of the stress tensor at the event horizon of the black string. We obtain the following results:
\begin{equation}\label{}
    T_{t}^{\ t}|_{horizon}=-\frac{3}{2}\frac{\alpha^{6}\eta}{\pi^{2}m^{2}}\left(\frac{1}{40}+3\eta^{2}\right)+\frac{\alpha^{6}}{140\pi^{2}m^{2}}
\end{equation}
\begin{equation}\label{}
    T_{z}^{\ z}|_{horizon}=T_{\varphi}^{\ \varphi}|_{horizon}=-\frac{3}{2}\frac{\alpha^{6}\eta}{\pi^{2}m^{2}}\left(\frac{1}{20}+3\eta^{2}\right)+\frac{\alpha^{6}}{112\pi^{2}m^{2}}
\end{equation}
\begin{equation}\label{}
    T_{\rho}^{\ \rho}|_{horizon}=-\frac{3}{2}\frac{\alpha^{6}\eta}{\pi^{2}m^{2}}\left(\frac{1}{40}+3\eta^{2}\right)+\frac{\alpha^{6}}{140\pi^{2}m^{2}}
\end{equation}
In the general case, all the components of the renormalized stress energy tensor of the quantized scalar field will be possitive at the horizon for the values of the coupling constant satisfying the
relation:
\begin{equation}\label{}
    3\eta^{3}+\frac{1}{40}\eta < \frac{1}{210}
\end{equation}
There are some particular cases in which the above relation is always satisfied. The simplest case of the conformal coupling \(\xi=\frac{1}{6}\) and the minimal one are two important examples. Also
for the case \(\xi<\frac{1}{6}\) the spacetime components of the quantized scalar field at the horizon of the black string are always positive quantities. If we define the energy density as usual:
\begin{equation}\label{}
    \varepsilon=-T_{t}^{\ t}
\end{equation}
then we can conclude that for the particular cases mentioned above (satisfying condition 32) the weak energy condition is violated. However, violations of the weak energy condition for quantum
matter are common, as in the case of Casimir Effect, and it is unknown how relevant the classical energy conditions are for this cases. Also, they  are in fact required for a self consistent picture
of the Hawking evaporation effect.

%%%%%%%%%%%%%%%%%%%%%%%%%%%%%%%%%

\section{Concluding remarks}
The quantization of a massive  scalar field  with arbitrary coupling to a gravitational background corresponding to cylindrical symmetry was considered. The renormalized quantum mean values of the
Lagrangian and the corresponding components of the Energy-Momentum tensor are explicitly evaluated for arbitrary axis symmetric space-time metrics configurations up to the second order in the
inverse mass of the scalar field. It is found that, as is usual for the case of quantum fields in curved backgrounds, exist regions in which the weak energy condition is violated. In the case of the
Black String considered in this work, we proof that the violation occur at the horizon of the spacetime for values of the coupling constant satisfying the relation (32), that include as particular
cases the most interesting of minimal and conformal coupling. Also for values of the coupling constant \(\xi<\frac{1}{6}\) the violation occur. This work furnish the starting point for of an
iterative solution of the quantum back-reaction problems of the quantum dynamics of the matter fields on the space-time metric. Due to the homogeneity of the scalar field equation of motion, the
results turn to be also valid for the quantization over non-vanishing scalar mean field configurations.
  The general formula are here employed to  explicitly evaluate
compact expressions for the mean value of the quantum Lagrangian
and  Energy Momentum tensor for the scalar field quantized over
the metric associated to the Black String  solution.
  As mentioned above, the results are expected to be employed to
investigate the back-reaction of the quantum scalar field, on the
Black String metric. For this purpose, the Einstein  equations for
the metric should be solved after including in them the calculated
 Energy-Momentum tensor for the Black String solution.  After that,
 the new value of the metric can be employed to determine the
 modified  Energy-Momentum tensor by substituting in  (\ref{emTensor}).
 Then, these new back-reacted components can be employed
 to solve the Einstein equations with them included again, and so on... iteratively.
  We expect to implement this program in coming  works.
\begin{acknowledgements}
One of the authors (O.P.F.P) greatly acknowledges M. Chac\'{o}n Toledo and E. R. Bezerra de Mello for helpful discusions. This work was suported by the Abdus Salam International Centre for
Theoretical Physics, Trieste, Italy.
\end{acknowledgements}

\end{document}